
\input harvmac.tex
\def\thm{{\bf \S}}
\nref\citeBUM{R.J. Bumcrot, ``Modern Projective Geometry", Holt, Reinhart, and
Winston, Inc.
(1969).}
\nref\citeHALL{ M. Hall, ``Theory of Groups", MacMillan (1959).}
\nref\citeWEIN{ S. Weinberg, ``Gravitation and Cosmology", John Wiley and Sons
(1972).}
\nref\citeBELIN{J. Belinfante, J. Math. Phys. {\bf 17} No. 3, 285 (1976).}
\nref\citeMIEL{ B. Mielnik, Commun. Math. Phys. {\bf 9} 55 (1968).}
\nref\citeSIER{ W. Sierpinski, ``Introduction to General Topology", U. Toronto
Press (1934).}
\nref\citeBvN{ G.D. Birkhoff and J. von Neumann, Ann. Math. {\bf 37}, 823
(1936).}
\nref\citeVAR{ V.S. Varadarajan, ``Geometry of Quantum Theory", Van Nostrand,
Princeton, (1968).}
\nref\citeGLEA{A.M. Gleason, J. Math. \& Mech. {\bf 6} 885 (1957).}
\nref\citeJWvN{P. Jordan, J. von Neumann, and E. Wigner, Ann. Math. {\bf 35},
29 (1934).}
\nref\citeFIV2{D.I. Fivel, University of Maryland preprint, UMD 1993-132. hepth
9404178}
\nref\citeWOLF{J.A. Wolf, ``Spaces of Constant Curvature". Publish or Perish
Inc. (1984).}
\nref\citeTRIG{ L. Todhunter and J. Leathem, ``Spherical Trigonometry",
MacMillan (1960).}
\nref\citeWANG{ H.C. Wang, Ann. Math. {\bf 55} 177 (1952).}
\nref\citePON{ L.S. Pontriagin, ``Topological Groups", Gordon and Breach
(1966).}
\nref\citeFIV1{ D.I. Fivel, Phys. Rev. Lett. {\bf 67}, 285 (1991).}

\Title{\vbox{\baselineskip12pt\hbox{UMD-PP 94-133}}}
{\vbox{\centerline{How Interference Effects in Mixtures}
\vskip9pt
\centerline{ Determine the Rules of Quantum Mechanics }}}
\vskip9pt
\centerline{Daniel~I.~Fivel~\foot{fivel@umdhep.umd.edu}}

\bigskip\centerline{\it Department of Physics}
\centerline{\it University of Maryland}
\centerline{\it  College Park, MD 20742}

\vskip 13mm

\centerline{\bf Abstract}
\vskip 4mm
It is shown that elementary indistinguishability properties of partially
polarized mixtures are
consistent only with the conventional Hilbert space model of quantum mechanics
and a few exotic
alternatives. This applies even in low dimensions where quantum logic and
Gleason's
theorem give either weak or no constraints. Experimental methods for
eliminating the exotic cases
(which include quaternionic and octonionic variants of quantum mechanics) are
described.
\vskip 4mm
\Date{\hfill 5/94}
\vfill\eject

\centerline{{\bf I. STATEMENT OF THE PROBLEM.}}
\vskip.1in
There is no reason to doubt the validity of the conventional model of quantum
mechanics wherein
 pure states $x$ are identified with projectors $ \pi(x) = |x><x|$ in a complex
Hilbert space
that may be finite or infinite dimensional, and the attenuation of an $x$-beam
by a
$y$-filter is given by
$$
p(x|y) = Tr(\pi(x)\pi(y)) = |<x|y>|^2.
\eqno(1)$$
In spite of its success there seems to exist no
demonstration of the {\it necessity} of the conventional model, and so from
time to time there have
been speculations that other models might exist (possibly less abstract ones!)
that agree with it
to the extent that predictions have been verified. The value of such
speculations lies in the
possibility of identifying new and subtle effects distinguishing various
models. A
historic precedent is found in geometry: Euclid's axioms codified our phyisical
intuition about the
structure of a plane. Cartesian coordinates were subsequently introduced as a
model for the plane,
and this worked so well many people came to  think of the Euclidean plane as no
different than the
Cartesian plane. But then it was discovered ~\citeBUM~,~~\citeHALL~~ that
certain theorems, e.g.\ the
theorems of Desargues and of Pappus, are deducible in the Cartesian plane but
do {\it not} follow
from Euclid's axioms. When one investigates the reason one learns (among other
surprising things)
that while the Cartesian plane can be embedded in a higher dimensional space,
not all Euclidean
planes can! Thus the physical property of embeddability in a three space is
actually a  fundamental
piece of information about the plane as we experience it whose significance we
would not have
appreciated had we not inquired into the  necessity of the Cartesian model.

Analogously a demonstration that a certain elementary set of physical facts
{\it compel} us to adopt the conventional model of quantum mechanics rather
than any other will
reveal that certain ingredients of the conventional model (whose significance
we may have
overlooked) are quite essential in distinguishing that model from others.  The
purpose of this paper
is to provide such a demonstration.
\vskip.1in
\centerline{{\bf II. PRELIMINARIES.}}
\vskip.1in
We shall place ourselves in the position of an experimentalist who is
unpredjudiced as to what model will describe the data. This data will be a
table, to be called a
$p$-{\it table}, consisting of  measured  attenuations $0 \leq p(x|y) \leq 1$
as $x,y$ range over a
set ${\cal S}$ of filters. Our task is similar to that of inferring the
geometry of the earth from a
table of road distances between cities~\citeWEIN~, and we can make this analogy
quantitative by
noting that there is a remarkably simple way~\citeBELIN~ of ``metrizing" the
$p$-table as follows:
Let
$$
d_L(x,y) = \sup_{z\in {\cal S}}{|p(z,x) - p(z,y)|},
\quad d_R(x,y) = \sup_{z\in {\cal S}}{|p(x,z) - p(y,z)|},
$$
$$
d(x,y) = max\{d_L(x,y),\; d_R(x,y)\}.
\eqno(2)$$
Since we make no {\it a priori} assumptions about the underlying physical
structure of the filters
we recognize {\it equivalence} $x=y$ of filters from $p(x|z) = p(y|z)$ and
$p(z|x) = p(z|y)$
for all $z \in {\cal S}$. One then readily checks that $d$ has the three
properties required of a
{\it metric} i.e.\
$$
x=y \Leftrightarrow d(x,y) = 0,\quad d(x,y) = d(y,x)
$$
$$
d(x,y) + d(y,z) \geq d(x,z), \; \forall x,y,z \in {\cal S}.
\eqno(3)$$
The only assumption about $p$ used in establishing (3) is that it is real (in
fact one one needs
only that it belongs to a ``valuation ring''). Thus ${\cal S}$ has now become a
metric space under
$d$, which, because of its generality, we call the {\it universal metric}.

The {\it symmetry group} ${\cal G}$ of ${\cal S}$ is the group of its
permutations that preserve
$p$ i.e.\ maps $z \to \tilde{z}$ such that $p(x|y) = p(\tilde{x}|\tilde{y}),\;
\forall x,y \in
{\cal S}$. Study of the symmetry group ${\cal G}$ is an obvious tool for
analyzing the structure of
${\cal S}$. In the conventional model there is a symmetry, which may be called
{\it exchange
homogeneity}, that exchanges any given pair of filters. While this is simply
stated it is not so
easy to test experimentally. However, it has an elementary consequence that
{\it is} easy to test
namely that $p(x|x) = p(y|y)$ and $p(x|y) = p(y|x),\; \forall x,y \in {\cal
S}$. We shall take this
as our first assumption about the $p$-table. Experimentally one finds the
common value of
$p(x|x)$ to be unity.

 Thus we shall assume:
$$
p(x|x) = 1, \; p(x|y) = p(y|x) \;\; \forall x,y \in {\cal S},
\eqno(4)$$
whence we may rewrite $d$ in the simpler form:
$$
 d(x,y) = \sup_{z\in {\cal S}}{|p(x,z) - p(y,z)|}.
\eqno(5)$$
It is important to note that we {\it do not} have to {\it assume} the converse
of the first property
in (4) i.e\   that $p(x|y) =1$ implies $x =y$, for we shall in fact be able to
derive it
below. In this connection one may contrast our approach with Mielnik's
{}~\citeMIEL~ analysis of
quantum mechanics via the function $p$ where $p(x|y) = 1$ {\it defines}
equivalence of filters $x,y$
rather than our definition $d(x,y) = 0$.

{}From its definition and the restriction $0 \leq p \leq 1$ we see that $0 \leq
d \leq 1,$ and from
(4) that any set $x_1,x_2,\cdots $ such that $p(x_j|x_k) = 0$ for $j \neq k$
are mutually separated
by the maximal amount $d = 1$. We shall say that $x,y$ are ``orthogonal" if
$p(x|y) = 0$ and define
a ``basis" as any maximal set ${\cal T}$ of mutually orthogonal elements.
While we introduce these
terms because they correspond to analogous terms in the conventional model, it
is important to keep in mind that we have as yet  no vector space in our
construction,  and hence the
reader is warned  not to ascribe properties to the terms beyond their
definition.

 It is to be noted
that in classical physics, where probabilities become certainties,  $p$ only
assumes the values
$0,1,$ so all of ${\cal S}$ is a basis. More generally a basis ${\cal T}$
represents
a maximal subset of ${\cal S}$ that behaves classically. In the semi-classical
case it will be
possible to partition ${\cal S}$ into disjoint subsets that are mutually
orthogonal. We may imagine
that this process of ``reducing" ${\cal S}$ to the union of orthogonal
components is carried out
until it can no longer be done, and the final components are then said to be
{\it irreducible}.

Let us next give an argument showing that the inherent limitations of
experiment
 are such that we can assume without loss of generality that ${\cal S}$ is a
{\it compact} metric
space, i.e.\ that infinite sequences have limit points:

In any experiment there will be a parameter $\epsilon > 0$ indicating the range
of error,
and we can regard two models $p_1,p_2$ as ``$\epsilon$-{\it equivalent}" if
$|p_1(x|y) - p_2(x|y)| <
\epsilon,\; \forall x,y \in {\cal S}$. Moreover we must be able to decide the
$\epsilon$-equivalence of two models with a  finite number of measurements that
may of course
increase as $\epsilon$ becomes smaller. Now let $S(x,\epsilon)$ be the ``ball''
 consisting of points
$y \in {\cal S}$ such that $d(x,y) < \epsilon$.  Suppose we  construct an
$\epsilon$-``cover" with a
set of balls $S(x_j,\epsilon)$, $j=1,2,\cdots M$. The triangle inequality
implies that $|d(x,z) -
d(x_j,z)| < \epsilon, \; \forall z \in {\cal S}$.  Thus the reqirement that we
be
able to determine the accuracy of the model by a finite number of measurements
is implemented by
requiring that there be a finite cover for every $\epsilon$. The topological
term for this property
is ``total boundedness" ~\citeSIER~. Moreover from an experimental point of
view, a filter is
indistinguishable from a filter that is sufficiently close in the $d$-metric,
so we may also assume
that every descending sequence of closed balls with $\epsilon \to 0$
contains only one common point. The topological term for this is the
``condition of Ascola". It can
then be shown that the space is ``complete", i.e.\ that Cauchy sequences
converge. Moreover one can
show~\citeSIER~ that a metric space is compact if and only if it is complete
and totally
bounded. Thus to any desired degree of experimental accuracy our $p$-table can
be associated with
a compact metric space. This means that there is at most a finite number of
elements in any
maximally separated set and hence that any basis ${\cal T}$ must contain a
finite number of
elements. Note that we do not have to assume that this number is the same for
every basis as we
shall be able to deduce this below. In the conventional model our conclusion
here corresponds to
the the fact that to any {\it given} experimental accuracy the predictions made
using {\it infinite}
dimensional Hilbert spaces ---which are not even locally compact--- can be
approximated by
restricting to a compact subspace i.e.\ a  Hilbert  space of sufficiently large
but {\it finite}
dimension.

As we noted earlier we expect that the mathematical
structure of ${\cal S}$ will be determined by its symmetry group ${\cal G}$.
In addition to the exchange homogeneity, of which (4) is a consequence, the
conventional model
enjoys two others worth noting at this point, for they will play an important
role in our
discussion below:

Consider the subgroup ${\cal
G}({\cal T})$ of ${\cal G}$ that fixes every element of a basis ${\cal T}$. In
the conventional
model these are unitary transformations obtained by exponentiating hermitian
operators for which
the states projected by the elements of ${\cal T}$ are an eigen-basis. Since
such operators commute
one sees that ${\cal G}({\cal T})$ is a {\it commutative} group. This property
of the conventional
model is the one used to argue that integrals of the motion commute with the
hamiltonian. We shall
see at the end of the paper that it is precisely this property that
distinguishes the conventional
model from certain exotic possibilities that are consistent with the other
assumed
properties of the $p$-table.

Next we note that the conventional model enjoys a symmetry property known as
{\it pairwise
homogeneity}, i.e.\ given two pairs of elements $a,b$ and $x,y$ such that
$p(a|b) = p(x|y)$, then
there exists a map in ${\cal G}$ that takes $a$ to $x$ and $b$ to $y$. This
property has a
remarkable consequence: In general (5) does not determine $p(x|y)$ from
$d(x,y).$
However if pairwise homogeneity holds one sees that if $p(a|b) = p(x|y)$ one
can use the
$p$-preserving mapping $a \to x$ and $b \to y$ to argue that the right sides of
$(5)$ will be the
same so that $d(a,b) = d(x,y)$. Thus there will exist some functional
relationship:
$$
d(x,y) = f(p(x|y)).
\eqno(6)$$
Since $f$ is also monotone in the conventional model (see Appendix), its
inverse exists and the {\it
isometry group} with respect to $d$ can be identified with the symmetry group
${\cal G}$.

While it is tempting to assume the pairwise homogeneity property we shall not
do so for the same
reason that we avoided assuming exhchange homogeneity above, namely that it is
not a propertly that
lends itself to easy experimental test. Rather we shall focus on the
relationship (6), seeking an
easily verified phenomenon that determines the form of (6). Remarkably we will
find that this
phenomenon determines all that we could have extracted from an assumption of
pairwise homogeneity.
\vskip.1in
\centerline{{\bf III. INDISTINGUISHABILITY PROPERTIES OF MIXTURES.}}
\vskip.1in
 Since {\it interference} is
the hallmark of quantum mechanics, one suspects that the sought-after phenomena
must be of this kind.
However,  we have the problem that interference is normally expressed in terms
of phase relations in
superpositions of states, and the notion of ``phase" has no meaning prior to
the formulation of a
complex vector space model. Thus we must first recognize those interference
phenomena that can be
expressed in a model-independent way i.e.\ {\it directly in terms of  entries
in the p-table.} Such
relations can be obtained from the study of {\it mixtures}.  A mixture ${\cal
M}$
consisting of  a fraction $\alpha_j$ of particles in the state $x_j$, $0\leq
\alpha_j \leq 1$,  $j =
1,2,\cdots, N$ will be denoted:
 $$
{\cal M} = \sum_j{\alpha_j x_j},\qquad \sum_j{\alpha_j} = 1.
\eqno(7)$$
It must be understood clearly that prior to the construction of a model this is
merely a formal
shorthand since the notion of linear combination of filters is not defined.
However, the fraction
of  ${\cal M}$ that passes a $z$ filter {\it is} well-defined, i.e.\
$$
P({\cal M}|z) = \sum_j{\alpha_j p(x_j|z)},
\eqno(8)$$
so that equivalence of filters can be defined by:
$$
{\cal M} = {\cal M}' \leftrightarrow P({\cal M}|z) = P({\cal M}'|z)\quad
\forall z\in {\cal
S}.
 \eqno(9)$$
{\it The kind of interference pheomena we are seeking will be expressed as
statements of equivalence
between mixtures constructed in different ways.} Unlike interference of
amplitudes they are
formulated directly from data in the $p$-table without benefit of a vector
space model.
In the Hilbert space formalism interferences of mixtures occur because
off-diagonal elements of
density matrices are in general complex numbers and so may cancel upon
addition.

 Using any basis  $\{x\}_N \equiv x_1,x_2,\cdots, x_N$ we can construct an {\it
unpolarized}
mixture:
 $$
{\cal U}(\{x\}_N) \equiv \sum_{j=1}^N{ N^{-1}x_j},
\eqno(10)$$
containing equal fractions of each of the basis states. Then the first of the
experimentally
verifiable interferences of probability that we are going to assume is quite
familiar:  {\it All
unpolarized mixtures are equivalent}. Thus, e.g.\ one cannot distinguish an
equal mixture of left
and right circularly polarized light from an equal mixture of orthogonal
linearly polarized states.
Thus we assume:
 $$
{\cal U}(\{x\}_N) = {\cal U}(\{y\}_{N'}) \hbox{ for any bases }
\{x\}_N,\;\{y\}_{N'},
$$
or equivalently
$$
N^{-1}\sum_{j=1}^{N}{p(x_j|z)} = N'^{-1}\sum_{k=1}^{N'}{p(y_k|z)}\qquad \forall
z\in {\cal S}.
 \eqno(11)$$

We will see below that (4) implies $N=N'$. In the conventional model  (11)
follows from the fact that
the density matrix for  unpolarized mixtures is a multiple of the unit matrix
and hence the same in
any basis. But (11) is  {\it also} consistent with  various hidden variable
models, and thus
 will {\it not} suffice to distinguish the conventional model. However we
observe that there
is another elementary indistinguishability property that can be observed using
{\it partially}
polarized mixtures. There are two simple ways to make such mixtures: by taking
an equal mixture of
non-orthogonal filters or an unequal mixture of orthogonal filters. It then
turns out that for any
mixture of the one type there is an equivalent mixture of the other. More
precisely: Given any $a,b$
there is an orthogonal pair $c,c'$ and a number
 $0 \leq \lambda \leq 1$ such that:
$$
\textstyle{1 \over 2}a +\textstyle{1 \over 2}b =  \lambda c + (1 - \lambda) c',
$$
or equivalently
$$
\textstyle{1 \over 2} p(a|z)  + \textstyle{1 \over 2} p(b|z) =  \lambda  p(c|z)
+ (1 -
\lambda )p(c'|z), \quad \forall z \in {\cal S}.
\eqno(12)$$
In the conventional model one deduces (12)  by diagonalizing the density matrix
associated with the
left side obtaining a result with non-vanishing elements in a two-dimensional
subspace associated
with the right side. From our point of view it is a simply testable property
described in a model
independent way that, as we shall see, contains almost all that we need to
deduce the conventional
model.
\vskip.1in
\centerline{{\bf IV. STATEMENT OF THE MAIN THEOREM .}}
\vskip.1in
We now summarize the results of our discussion above and formulate the main
theorem to be
proved below:

We assume that a function $0 \leq p(x|y) \leq 1$ called an {\it attenuation
function} is given on
pairs of elements in a set called the {\it space of filters}. This becomes a
metric space under the
universal metric $d$ determined by $p$ (see (2)). Any  compact subspace (with
respect
to $d$) can be decomposed into the finite union of subspaces
that are mutually orthogonal in the sense that $p(x|y) = 0$ when $x$
and $y$ belong to different subspaces and are irreducible in the sense that
they cannot be further
decomposed in this way. We refer to these compact, irreducible subsets as {\it
components}. A
typical component is denoted ${\cal S}$. A maximal set ${\cal T}$ of mutually
orthogonal elements of ${\cal S}$ is called a {\it basis} of ${\cal S},$ and it
follows from
compactness that the number of elements in any such set is finite so that it
makes sense to define
unpolarized mixtures by $(10).$

In the following ${\cal R},\; {\cal C},\;{\cal Q}$ and ${\cal C}ay$ refer to
the real numbers,
complex numbers, quaternions, and octonions (Cayley numbers) respectively.

The main theorem tells us the structure of ${\cal S}$ subject only to the
following assumptions
about the attenuation function:

(i)  $p(x|x) = 1,\;\; p(x|y) = p(y|x), \; \forall x,y \in {\cal S}.$

(ii) Unpolarized mixtures are indistinguishable  $(11).$

(iii)  The averaging property $(12)$ of partially polarized mixtures holds.

\vskip.1in
\centerline{\thm {\bf (4.1)} {{\bf {\it Main Theorem}}}}
\vskip.1in
\thm({\bf A}) {\it All bases have the same number of elements} $N$ {\it called
the ``dimension" of} ${\cal S}$.

\noindent
\thm ({\bf B}) {\it The attenuation function} $p$ {\it is related to the
universal metric by the
 formula:}
$$
d(x,y) = \{1 - p(x|y)\}^{1/2}.
$$
{\it In particular this means that isometries with respect to } $d$ {\it are
symmetries of} ${\cal
S}$.

\noindent
\thm({\bf C}) {\it For} $N=2$ {\it there is an isometry of} ${\cal S}$ {\it to
a unit sphere of
finite dimension} $m$ {\it on which the metric is the great circle arc-length.
The two elements of a
basis lie at antipodes, and}
 $p(x|y) = \cos^2(\theta(x,y)/2),$ {\it where }$\theta$ {\it is the great
circle arc-length between
the corresponding points. In the cases} $m=1,2,4,8$ {\it  the sphere is also a
projective space
over} ${\cal R},\; {\cal C},\;{\cal Q}${\it or} ${\cal C}ay,$ {\it respectively
and the trace
rule (1) for}  $p$ {\it  holds.}

\noindent
\thm({\bf D}) {\it  For} $N = 3$ {\it the set } ${\cal S}$ {\it can be mapped
to a projective space
over}  ${\cal R},\; {\cal C},\; {\cal Q}$ {\it or} ${\cal C}ay,$ {\it in such a
way that the trace
rule (1) for} $p$ {\it holds.}

\noindent
\thm({\bf E}) {\it  For} $N > 3$ {\it the result is as for} $N = 3$ {\it except
that} ${\cal C}ay$
{\it is excluded.}

\noindent
\thm({\bf F}) {\it The conventional model of quantum mechanics (which
corresponds to} ${\cal R}$ {\it
or} ${\cal C}$ {\it projective spaces above) is the unique model for which the
subgroup of isometries
fixing the elements of a basis is commutative.}

In the course of the proof we shall indicate how our main theorem relates to
 the quantum logic of Birkhoff  and von
Neumann~\citeBvN~,~\citeVAR~, to Gleason's theorem~\citeGLEA~,  and to the
Jordan algebra
axiomatization scheme~\citeJWvN~. The physical significance of \thm{\bf (F)} of
the
main theorem will be discussed at the end of the paper.

The proof of \thm{\bf (A)} is quite simple: Insert $z= y_l$, and sum both sides
of (11) over $l$;
insert $z=x_n$, sum both sides over $n$, interchange, and use (4)  to deduce
that
$N = N'$. We also have the important corollary:

\thm {\bf (4.2)}  $p$ {\it is a ``frame function" ~\citeGLEA~ i.e.\ }:
$$
\sum_{j=1}^N{p(x_j,|z)}=1, \quad \forall z, \quad \hbox{ for any basis }
\{x\}_N,
\eqno(13)$$
To see this observe that any $z$ can be made part of a basis by adjoining
orthogonal elements to it
until a maximal set is obtained. The assertion then follows from (11). Note
that (13) is
the familiar assertion in quantum mechanics that the sum of the probabilities
is unity for a
particle to pass the filters belonging to a basis.

Proofs of the remaining parts of the main theorem are more complicated and
appear in various
sections below.

\vskip.1in
\centerline{{\bf V. DEDUCTION OF THE RELATION (B) OF $p$ TO THE METRIC.}}
\vskip.1in

Two lemmas will be needed: Let $n \leq N$ where $N$ is the dimension of the
system and let $\{v\}_n = v_1,v_2,\cdots,v_n$ be a mutually orthogonal set.
Then we define the {\it
subspace} ${\cal S}^*$ of ${\cal S}$ {\it spanned by} $\{v\}_n$ as the set of
states $u$ such that:
$$
\sum_{j=1}^{n}{p(v_j|u)} = 1.
\eqno(14)$$
The use of the term ``subspace" is justified by the following lemma:

\thm {\bf (5.1)} {\it If} $\{w\}_n$ {\it is any other mutually orthogonal set
of n elements in}
${\cal S}^*,$ {\it then it also spans} ${\cal S}^*$. {\it Moreover}
$$
\sum_{j=1}^{n}{p(w_j|z)} = \sum_{j=1}^{n}{p(v_j|z)}\qquad \forall z \in {\cal
S},
\eqno(15)$$
{\it the common value being unity if} $z \in {\cal S}^*$.

Proof: Let $\{v\}_n$ be extended to a basis of ${\cal S}$ by adjoining
orthogonal elements $v_{n+1},\cdots v_N$. Then from the definition and (13),
$p(v_j|u) = 0$ for
 $u \in {\cal S}^*$ and $j > n$. Hence $p(w_k|v_j) = 0$ for $j > n$ and hence
$w_1,\cdots,w_n,v_{n+1},\cdots v_N$ is a basis of ${\cal S}$. Comparing (13)
for these two bases the
assertion follows.$\bullet$

 Note that \thm {\bf (5.1)} shows that it
makes sense to speak of the dimension of a subspace as the (unique) number of
elements in any
maximal mutually orthogonal set in the subspace. Another useful corollary is
the following:
For given $x$ consider the one dimensional subspace consisting of the set of
elements $z$ such that
$p(x|z) = 1$. Since it is one-dimensional any element $y$ for which $p(x|y) =
1$ also spans
it, and hence, by \thm{\bf(5.1)}, $p(x|z) = p(y|z)$ for all $z \in {\cal S}$,
i.e.\  $x = y$. Combining this with (4) we have proved:
$$
x = y \leftrightarrow p(x|y) = 1.
\eqno(16)$$
We thus deduce the equivalence criterion assumed by Mielnik ~\citeMIEL~ noted
above. With (5) we also
obtain the converse of a relationship noted above between bases and maximally
separated
elements:

\thm {\bf (5.2)}: {\it Every maximally separated set of elements (in the sense
of the d-metric) is a
basis and vice versa.}

We have not yet used (12) but will now do so in proving:

\thm {\bf (5.3)} {\it If} $a \neq b$ {\it there is a unique two-dimensional
subspace }${\cal P}_{ab}$
{\it containing} $a,b$ {\it  as well as} $c,c'$ {\it appearing in} $(12)$.

Proof : Let $c_1 = c$, $c_2 = c'$ and extend  to a basis of ${\cal S}$ by
adjoining $N-2$ orthogonal  elements  $c_3,\cdots,c_N$. With $z = c_i, i > 2$
the right side of (12)
is zero and since the quantities on the left are non-negative, we must have
$p(a|c_i) = p(b|c_i) =
0$ for $i > 2$. Hence, by (13,14), $a,b$ are in the two-dimensional subspace
spanned by $c,c'$ Now
if $a,b$ are distinct $\lambda \neq 1.$  For otherwise $p(a|c') = p(b|c') = 0$,
and so $a,b$
lie in the one-dimensional subspace spanned by $c$, i.e.\ $a=b$. By similar
argument $\lambda
\neq 0$.  Now suppose $a,b$ belong to a subspace spanned by some other
orthogonal pair $d,d'$. Then
inserting $z = d$ and $z = d'$ into (12) and adding one has: $1 =
\lambda(p(c|d) + p(c|d') )+
(1-\lambda)(p(c'|d) + p(c'|d'))$. Since $0 < \lambda < 1$ this can be satisfied
only if
$p(c|d) + p(c|d') = p(c'|d) + p(c'|d') = 1,$ which means that $d,d'$ and $c,c'$
define the same
subspace.$\bullet$

{}From \thm {\bf (5.3)} we can make the following definition: If $b \neq a$
then the {\it antipode}
$a'$ of $a$ relative to $b$ is the unique element in ${\cal P}_{ab}$ such that
$p(a|a') = 0$.
Moreover since $a,a'$ and $b,b'$ span ${\cal P}_{ab}$ as well as $c,c'$ in
$(12)$ we have proved
that if $a \neq b$ then:
$$
p(a|z) + p(a'|z) = p(b|z) + p(b'|z)= p(c|z) + p(c'|z), \quad \forall z\in {\cal
S},
\eqno(17)$$
the common value being unity if $z \in {\cal P}_{ab}.$

For later use we take note of three corollaries:

\thm {\bf (5.4)}: ${\cal P}_{ab}$ {\it is determined by any pair of its
distinct elements. Thus if}
${\cal P}_1$ {\it and} ${\cal P}_2$ {\it are two-dimensional subspaces,  they
are either disjoint,
identical or have just one common point.}

 This will be exploited later on in demonstrating that our space has
among other things the structure of a {\it projective geometry}. Note: We shall
say that a pair of
two-dimensional subspaces are {\it adjacent} if they have exactly one common
point.

\thm {\bf (5.5)}: {\it If} $z$ {\it is orthogonal to two distinct points of a
two dimensional
subspace it is orthogonal to every point of that subspace.}

We say that a pair of adjacent two-dimensional subspaces are {\it normal} to
one another if the
antipodes of the intersection in the two subspaces are orthogonal to one
another.

\thm {\bf (5.6)}: {\it If} $z$ {\it is the intersection of a pair of normal
two-dimensional subspaces
then its antipode} $z'$ {\it on one of the two subspaces is orthogonal to every
point of the
other.}

We are now ready to prove ({\bf B}) of the Main Theorem \thm {\bf (4.1)}.

 If $a = b$, $p(a|b) = 1$ and there is nothing to prove. If $a \neq b$ then let
$b'$ be the
antipode of $b$ relative to $a$. If $b' = a$, so that $p(a|b)
= 0$ we already  know that $d(a,b) = 1,$ and ({\bf B}) follows.
Thus we may assume that $a \neq b,b'$.  Equations $(12,17)$ can be combined to
give:
 $$
p(a|z) - p(b'|z) = (2\lambda - 1)(p(c|z) - p(c'|z)) \; \forall z \in {\cal S}.
\eqno(18)$$
Taking the supremum over $z$, the definition (5) of $d$ gives:
$$
d(a,b') = |2\lambda - 1|d(c,c') = |2\lambda - 1|,
\eqno(19)$$
whence, since we assume $a \neq b',$ it follows that $\lambda \neq 1/2$.
{}From $(17)$ when $z \in {\cal P}_{ab}$ so that the common value is unity one
obtains:
$$
p(a|z) - p(b'|z) = (2\lambda - 1)(2p(c|z) - 1) = p(b|z) - p(a'|z).
\eqno(20)$$
In particular for $z = a$ and $z = b$:
$$
p(a|b) = (2\lambda - 1)(2p(c|a) - 1) = (2\lambda - 1)(2p(c|b) - 1),
\eqno(21)$$
so that since $\lambda \neq 1/2$:
$$
p(c|a) = p(c|b).
\eqno(22)$$
Then from (17) again with $z = c$
$$
\lambda = p(c|a),
\eqno(23)$$
so that from (21)
$$
|2\lambda - 1| = \sqrt{p(a|b)} = \sqrt{1 - p(a|b')},
\eqno(24)$$
which with $(19)$ and the substitution $x = a, y = b'$ gives
$$
d(x,y) = \{1 - p(x|y)\}^{1/2},
\eqno(25)$$
which is ({\bf B}) of \thm {\bf (4.1)}.$\bullet$

Since $d$ uniquely determines $p$ from  $p = 1 - d^2$ it also follows that
every isometry with
respect to $d$ is also a symmetry of ${\cal S}$ as defined above.

  The proof that
$(25)$ agrees with the conventional model is given in the Appendix where the
proof~\citeFIV2~ is
also reproduced that this relationship is distinct from the one obtained in
locally realistic (hidden
variable) theories. Thus it follows that although (11) can be reproduced by
locally realistic
theories, the relation (12) cannot!
\vskip.1in
\centerline{{\bf VI. CONVEXITY .}}
\vskip.1in
We next use (12) and the fact that ${\cal S}$ is irreducible to establish a
basic convexity property
of ${\cal S}$ that will be needed in the remaining parts of the proof of the
main theorem.

Let us examine the {\it uniqueness} of $\lambda,c,c'$ in (12) for given $a,b$.
Assume
that $p(a|b) \neq 0,1$. One notes that (24) has  exactly two solutions for
$\lambda$, one in the
interval $1 > \lambda > 1/2$ and the other its image under $\lambda \to 1 -
\lambda$. The exchange
of these two is equivalent to exhchanging $c$ and $c'$ in $(12)$.
If we fix $\lambda$ then the $c$ satisfying $(12)$ is unique for the following
reason: Since \thm
{\bf (5.1)} implies $c \in {\cal P}_{ab},$  \thm {\bf (5.3)}implies that for
any $z \in {\cal
P}_{ab}$ we must have $p(c'|z) = 1  - p(c|z)$ and hence from (12):

$$
 \textstyle{1 \over 2} p(a|z)  + \textstyle{1 \over 2} p(b|z) =  (2\lambda - 1)
p(c|z) + (1 - \lambda ) \quad \forall z \in {\cal P}_{ab}.
\eqno(26)$$

Hence if $\lambda > 1/2$ is given it follows that if $c_1$ and $c_2$ are two
different choices
for $c \in {\cal P}_{ab},$ then $p(c_1|z) = p(c_2|z)$ for all $z \in {\cal
P}_{ab}$. In particular
with $z = c_1$ we obtain $p(c_1|c_2) = 1$ so that $c_1 = c_2$ by (16). This
establishes the
uniqueness of $c$ for given $\lambda > 1/2,$ and we can give it the following
geometric
interpretation. Define the {\it
arc} $\theta(a,b)$ through the equation:
$$
p(a|b) = \cos^2(\theta(a,b)/2), \quad 0 \leq \theta \leq \pi,
\eqno(27)$$
so that
$$
\lambda = \cos^2(\theta(a,b)/4)
\eqno(28)$$
\noindent
is the solution of (24) with $\lambda \geq 1/2$. Then from (22,23) we have
$$
\theta(a,c) =
\theta(c,b) = \theta(a,b)/2.
\eqno(29)$$
We have thus established:

\thm {\bf (6.1)} {\it If} $p(a|b) \neq 0,1$ {\it there is a unique point} $c$
{\it in }
${\cal P}_{ab}$ {\it that may be called the midpoint of } $a$ {\it and} $b$
{\it satisfying }
$\theta(a,c) = \theta(c,b) = \theta(a,b)/2.$ {\it In other words} ${\cal S}$
{\it is an
``M-convex" metric space.} ~\citeWOLF~

In order that \thm {\bf (6.1)} be interesting it is necessary that there exist
$a,b$ for which
$p(a,b) \neq 0,1$. But this follows  from the irreducibility of ${\cal S}$.  We
will see below
that it also implies that ${\cal S}$ is {\it connected}.
\vskip.1in
\centerline{{\bf VII. PROOF OF \thm (4.1C) AND THE  STRUCTURE OF TWO
DIMENSIONAL SUBSPACES.}}
\vskip.1in
First we must caution the reader about the use of the term
``dimension" in the following discussion. Recall that the dimension of a
subspace ${\cal S}^*$ of
${\cal S}$ is the number of mutually orthogonal elements (as defined by $p$) or
equivalently the
maximal number of points with maximal separation. One should not confuse this
with the dimension
$m$ of a space {\it as a manifold}. Thus e.g.\ an $m$-sphere has dimension  $m$
as a manifold
but has exactly {\it two} maximally separated elements for any $m \geq 1$. In
particular
linearly polarized photon states are represented by points on a circle
(1-sphere) whereas the set of
all polarization states corresponds to the points of the Poincar\'{e} sphere (a
2-sphere). In both
cases there are exactly {\it two} orthogonal states in any basis. Since
linearly polarized states are
described by a real two-dimensional Hilbert space, whereas the set of all
polarization states
requires a complex two-dimensional Hilbert space, one sees that the increase in
manifold dimension is
associated with the enlargement of the coefficient field required, not in the
number of orthogonal
states.

 We shall say that a subset ${\cal P}^*$ of ${\cal P}$ is {\it properly mapped}
to a unit
$m$-sphere if for every pair of points $a,b$ in ${\cal P}^*$, the great-circle
arc $\theta(a,b)$
joining the corresponding points  satisfies $p(a|b) = \cos^2(\theta(a,b)/2).$
If some
 ${\cal P}^*$ is properly mapped to a unit m-sphere, then one can adjoin the
antipodes of
all elements, and the extended set is still properly mapped to the $m$-sphere.
This follows
simply from the fact that the antipode $a'$ of $a$ will, in virtue of $p(a|z) +
p(a'|z) = 1,\;
\forall z\in {\cal P},$ give the correct arc length to $\theta(a',z)$ whenever
$\theta(a,z)$ is
correct. We now show that ${\cal P}^*$ can be enlarged so that together with
any pair of its
points it also contains the {\it midpoint}: First rewrite (26) in terms of arcs
with
$\alpha,\beta,\gamma,\theta$ being the arcs of $p(a|z),p(b|z),p(c|z),$ and
$p(a|b)$ respectively
(see Figure 1). After some trigonometric manipulation it becomes:
 $$
\textstyle{1 \over 2}(\cos \alpha + \cos \beta) = \cos \gamma \cos(\theta/2).
\eqno(30)$$

This equation has a remarkable significance:   Suppose that it is possible to
put $z$,$a$,$b$
on a two-sphere in such a way that the great-circle arc lengths between them
agree with the
$\alpha,\beta,\theta$ defined above. In sphereical trigonometry it is shown
{}~\citeTRIG~ that (30) is
 the formula giving the length $\gamma$ of the
great-circle arc joining $z$ to the mid-point of the arc connecting  $a,b$.
Thus it would follow
that $c$ of $(12)$ is properly mapped to that {\it same} 2-sphere! But if three
points can be
properly mapped to an $m$-sphere they also belong to a $2$-sphere (possibly
degenerating to a 0 or 1
sphere) {\it within} it. Thus starting with any subset ${\cal P}^*$ we can
continue to adjoin
midpoints of all pairs, antipodes of all elements, and finally take the closure
in the metric
topology to form a set  $[{\cal P}^*]$ called the {\it m-closure} that is
closed with respect to all
three operations. Thus we have:

\thm {\bf (7.1)}: {\it If} ${\cal P}^* $ {\it can be properly mapped to an
m-sphere then so also can
its} $m${\it -closure} $[{\cal P}^*].$

{}From the assumption of irreducibility it follows that  given any element $x$
in ${\cal P}$ there must
exist in addition to its antipode $x'$ at least one  element $y$ such that $0 <
p(x|y) < 1$.
But clearly the set consisting of the three points $x,y,x'$ can be mapped to a
unit-circle with
$x,x'$ at opposite points of a diameter. By \thm {\bf (7.1)} so also can its
$m$-closure. But one
sees from the construction of the $m$-closure that this is simply the unit
circle itself. Thus we
have proved:

\thm {\bf (7.2)}  ${\cal P}$ {\it is connected. In fact
 every pair of points} $a,b \in{\cal P}$ {\it can be connected by a circular
arc} $\theta(a,b)$
{\it in such a way that}  $d(a,b) = \sin(\theta(a,b)/2)$ {\it and so is just
one-half the length of
the chord of the circular arc connecting the two points.}

We next introduce a useful tool for analyzing the geometry: For any $x$ in
${\cal P}$ we define
the {\it equator ${\cal E}(x)$ opposite} $x$ as the set of points $c$ with the
the property
$\theta(x,c) = \theta(c,x') = \pi/2.$ From connectedness this set is not empty.
Clearly $p(x|c) =
1/2 \to p(c'|x) = 1/2$ so $c\in {\cal E}(x) \to c' \in {\cal E}(x)$. Also if
$a,b \in {\cal E}(x)$
are not one another's antipodes so that $\lambda \neq 1/2$ in (26), then
solving (26) with $p(a|x)
= p(b|x) = 1/2$ gives $p(c|x) = 1/2$. Thus

\thm {\bf (7.3)} {\it The midpoint of two points in} ${\cal E}(x)$ {\it is also
in} ${\cal E}(x)$.

 Thus ${\cal E}(x)$ is a {\it proper} subset of ${\cal P}$ (since $x,x' \notin
{\cal
E}(x)$) that is {\it closed with respect to inclusion of antipodes and
midpoints} just as ${\cal P}$
itself. If ${\cal E}(x)$ is non-empty we can select an arbitrary point say
$x_1$ and define the
equator ${\cal E}_1(x_1)$ opposite $x_1$ as the subset of ${\cal E}(x)$
consisting of points $y$
satisying $p(y|x_1) = 1/2$. We continue in this way to generate a sequence
$x_1,x_2,\cdots $. At
some finite $m$ we will encounter an $x_m$ for which the equator opposite is
empty. The reason $m$
must be finite is that every point in the sequence is separated from the others
by $d = 1/\sqrt{2}$
in the metric, and hence, by compactness, such a sequence can have an at most
finite number of elements. Quantum mechanics over the reals, complex numbers,
quaternions, and
octonions (Cayley numbers) would correspond to $m=1,2,4,8$ respectively. We
call the sequence
$x_1,\cdots,x_m$ an {\it equatorial decomposition}.

 Now we claim that ${\cal P}$ is the $m$-closure of the set ${\cal P}^*$
consisting of any point $x$ and the equator ${\cal E}(x)$. To see this note
that if $a \notin {\cal
P}^*$ then there is a unique circle containing $x,a,x'$ that intersects ${\cal
E}(x)$ in a unique
point $y,$  and $a$ is in the $m$-closure of any set containing
$y$ and $x$. Now if $x$ and ${\cal E}(x)$ can be properly mapped to a
$j$-sphere, then clearly ${\cal
P}^*$ can be properly mapped to a $j +1$-sphere. By \thm {\bf (7.1)} so can its
$m$-closure, and
hence so can ${\cal P}.$ Hence, using the equatorial decomposition, we obtain
part {\bf (C)} of the
main theorem \thm {\bf (4.1)} by induction.$\bullet$

We now know that a two-dimensional subspace ($N=2$) is an $m$-sphere ($m\geq
1$). We will
call it  a {\it generalized Poincar\'{e} sphere (GPS).} To avoid confusion we
refer to $m$ as the
{\it rank} of the GPS rather than a dimension. It is  the number of points in
the equatorial
decomposition. The term {\it Poincar\'{e} sphere} without the adjective
``generalized" is used to
specify the conventional model ($m=2$). Thus in the case of polarized light we
may represent the pole
by the projector of the state $(1,0)$ and the two points of an equatorial
decomposition can be taken to be the projectors of the states $(1,1)/\sqrt{2}$
and $(1,i)/\sqrt{2}$.
Thus a maximal set with mutual attenuation equal to $1/2$ has three elements.
If the description of  polarization  required quaternions ($N=2, m=4$) we could
take an
equatorial decomposition $(1,1)/\sqrt{2},\; (1,i)/\sqrt{2},\;
(1,j)/\sqrt{2},\;(1,k)/\sqrt{2}$ where
$i,j,k$ are the quaternionic units, and we would find that there are five
``polarization" filters with mutual attenuation equal to $1/2$ rather than
three. In the octonionic
case ($m=8$) we would replace $i,j,k$ by the seven octonionic units and have
$8$ filters with mutual
attenuation $1/2$. (For arbitrary $m$ the units are associated with  all
possible Jordan
algebras~\citeJWvN~ - see below).

 Let us note here that while  $N=2$ is conceptually simpler
than $N > 2$, it is the most troublesome case for various axiomatic schemes.
Thus, e.g.\ in the
so-called quantum-logic approach of Birkhoff and von Neumann
{}~\citeBvN~,~\citeVAR~, no information is
gotten about $N=2$ because the fundamental theorem of projective geometry that
they exploit gives
non-trivial constraints only for $N > 2$. Moreover even if one assumes that one
has a Hilbert space
for $N=2$, it is not possible to deduce (1) from (11) alone  using Gleason's
theorem~\citeGLEA~
(see discussion below).  It is thus encouraging that (12) has supplied us with
new structure not
present in quantum logic. We will see next that $(12)$ also gives us more for
the difficult case $N
= 3$ than one obtains from quantum logic.
\vskip.1in
\centerline{{\bf VIII.  DIMENSION $N = 3.$}}
\vskip.1in
We have already shown that the geodesic for $N=2$ is a great circle arc. We
want to show next that
in higher dimensions the great circle arc on the generalized Poincar\'{e}
sphere joining two points
is still the shortest of {\it  all} possible rectifiable curves joining them.

\thm {\bf (8.1)} {\it The geodesic
connecting any two filters is an arc of great circle lying in the generalized
Poincear\'{e} sphere
containing them.}

Proof: $d(a,b)$ is one-half the chord length of the circular arc between $a,b$
in ${\cal P}_{ab}.$ The great circle arc joining $a,b$ has length
$\theta(a,b)/2$ in this metric.
Hence if ${\cal F}$ is a rectifiable curve of shortest length $\phi(a,b)$ one
has:
$$
\sin(\theta(a,b)/2) \leq \phi(a,b) \leq \theta(a,b)/2.
\eqno(31)$$
Mark off  points $a = a_o, a_1, a_2, \cdots , a_n = b$ on the curve
 so that the segments connecting adjacent points have equal length $\phi/n$.
Then by (31) this
differs in length from the circular arc joining a pair of adjacent points by an
amount of order
$(\phi /n)^2$ and hence for the whole curve the error is of order $\phi^2/n \to
0, n \to \infty$.
Hence we can approximate ${\cal F}$ to arbitrary accuracy by a sufficiently
large number of
circular arcs joining adjacent points. We must establish that the shortest
length is obtained when
those arcs lie on the {\it same } circle. To do this we proceed as follows:

Recall that (26) required $z \in {\cal P}_{ab}$ because it was derived from
(12) using $p(c|z) +
p(c'|z) = 1.$ However for arbitrary $z$ one sees that $p(c|z) + p(c'|z) \leq
1$. Hence (26)
generalizes as an inequality for arbitrary $z$ and, when expressed in terms of
arcs, (30) becomes
in the special case $\alpha = \beta$:
 $$
\cos \alpha \leq \cos \gamma \cos(\theta/2) \leq \cos(\theta/2) \; i.e.\ \;
\alpha \geq \theta/2,
\eqno(32)$$
{\it in which the arcs are no longer required to lie on the same sphere.} But
this simply says that
the sum of two equal circular arcs connecting $a$ to $z$ and $z$ to $b$ is not
smaller
than the length of of a single great circle arc connecting $a$ to $b,$ which is
what we had to
prove.$\bullet$

\thm {\bf (8.2)} {\it All generalized Poincar\'{e} spheres in} ${\cal S}$ {\it
are of the same rank.
The rank} $m$ {\it of these spheres is called the rank of} ${\cal S}$.

Proof: From \thm {\bf (5.4)} two GPS are either disjoint, have one point in
common, or are identical. If they are disjoint there is a GPS containing one
point from each.
Hence it suffices to prove the theorem for a pair of adjacent GPS's. Connect
the antipodes of the
intersection point $x$  by a great circle arc  and divide this up into a large
number of
equal segments. Then we have a chain of closely spaced points generically
denoted $y$.
Clearly each such $y$ satisfies $p(y|x) = 0$ and hence is also the antipode of
$x$ on some GPS
denoted ${\cal P}_{xy}$. Thus we need only show that for  GPS's with closely
spaced $y$ the dimension
will be the same, i.e. that the number of points in the equatorial
decomposition cannot suddenly
jump for ${\cal P}_{xy}$'s with nearby $y$'s. But since $|d(a,x_1) - d(a,x_2)|
\to 0$ for
$d(x_1,x_2)$ by the triangle inequality it follows that if $d(a,x_1) = 1/2$
then $d(a,x_2)$ can be
made arbitrarily close to 1/2 for $x_1$ and $x_2$ close enough. Hence the
sequence of ${\cal
P}_{xy}$'s have an equatorial decomposition retaining the same number of points
in the limit $y \to
x_1$ $\bullet$.

If $z$ is a point not on the GPS ${\cal P}$ we shall define a {\it foot} of $z$
on ${\cal P}$ as
a point $f$ of ${\cal P}$ for which $d(z,f)$ is a minimum. We shall see that if
$z$ is not
orthogonal to ${\cal P}$ then its foot is unique. Moreover we shall deduce an
analogue of the
Pythagorean theorem namely:

\thm {\bf (8.3)} {\it If} $x \in {\cal P}$ {\it then} $p(z,x) =
p(z,z^*)p(z^*,x)$ {\it where} $z^*$
{\it is the foot of} $z$ {\it in} ${\cal P}$.

To construct the foot $z^*$ in ${\cal P}$ of $z \notin {\cal P}$ and not
orthogonal to ${\cal P}$
proceed as follows (see Figure 2):  Select an arbitrary pair of antipodes
$a,a'$ of ${\cal P}$ and
let $a^*$ be the antipode of $a$ in ${\cal P}_{az}$. This will be distinct from
$a'$ since $z \notin
{\cal P}$. Let $a''$ be the antipode of $a'$ in ${\cal Q} \equiv {\cal
P}_{a'a^*}$ and $a^{**}$ be
the antipode of $a^*$ in ${\cal Q}$. Since $a$ is orthogonal to $a'$ and $a^*$
it is by \thm {\bf
(5.5)} orthogonal to ${\cal Q}$ and hence $a,a',a''$ are mutually orthogonal.
Also $a,a^*,a^{**}$ are
mutually orthogonal and so, since $p(a|z) + p(a^*|z) = 1$ it follows that
$p(z|a^{**}) = 0$. But
$p(a^*|z) + p(a^{**}|z) = p(a'|z) + p(a''|z)$ by \thm {\bf (5.1)} and so
$p(a|z) + p(a'|z) + p(a''|z)
= 1$. Hence $z$ is in the subspace spanned by $a,a',a''$. One notes that since
$z$ is by hypothesis
not orthogonal to ${\cal P}$ it will be different from $a''$, and hence  the
GPS ${\cal P}_{a''z}$
will intersect ${\cal P}$ at a unique point. We take this to be the {\it foot}
$z^*$ of $z$ on ${\cal
P}$. Note that  ${\cal P}$ and ${\cal P}_{a''z}$ are {\it normal} to one
another as defined
following \thm {\bf(5.5)}. Thus \thm {\bf (8.3)} can be regarded as a formula
relating distances
between two points on a pair of normal GPS's to their distances to the
intersection point, in this
case at $z^*$.

To prove \thm {\bf (8.3)} we need two lemmas:

Let $z_1,z_2$ lie respectively on a normal pair of GPS's ${\cal P}_1$ and
${\cal P}_2$ with
intersection $z_o$.

\thm {\bf (8.4)}  $z_o$ {\it is the nearest point of} ${\cal P}_2$ {\it to any
point of} ${\cal
P}_1$.

To see this observe that if there were a point $x$ on ${\cal P}_2$ nearer to a
point $y$ on
${\cal P}_1$ and if $z_o'$ is the antipode of $z_o$ on ${\cal P}_1,$ then the
curve consisisting of a
circular arc connecting $z_o'$ to $y$ together with the circular arc connecting
$y$ to $x$ would be
less than a semicircle. But they can be connected by a
semi-circular arc which by \thm {\bf (8.1)} is the shortest distance between
them.$\bullet$

{}From (18,28) we have
$$
p(a|z) - p(b'|z) = -\sin(\theta(a,b')/2)(p(c|z) - p(c'|z)).
\eqno(33)$$
Now let the intersection $z_o$ of the two normal spheres be the midpoint of the
two points $a,b'$ on
${\cal P}_2$. Consider all pairs of points  on the circular arc ${\cal C}$
joining $a,b'$
that are equidistant from $z_o$. Replacing $a,b'$ by these will not change
$c,c'$ which will always
lie at opposite poles one quarter circle from $z_o$ on ${\cal C}$. Now let
 $x$ be a variable point on the circular arc joining $a,b'.$  Then for any $y$
on ${\cal P}_1$
\thm {\bf (8.1)} shows that $p(x|z)$ is a minimum when $x$ passes through
$z_o$. Hence the left
side of the last equation must vanish faster than linearly in $\theta (a,b')$
as this quantity tends
to zero. But the first factor on the right of (33) vanishes linearly so that we
must have $p(c|z) -
p(c'|z) = 0$ and hence $p(a|z) = p(b'|z)$. In other words we have proved the
lemma:

\thm {\bf (8.5)} {\it If} ${\cal P}_1,{\cal P}_2$ {\it are normal, two points
of} ${\cal P}_1$
{\it equidistant from the intersection are also equidistant from any point of}
${\cal P}_2$.

Now if $a,b$ in (12) are points of ${\cal P}_2$ equidistant from the
intersection $z_o$, then in
(12) we see that $c=z_o.$ Moreover $c'$ is its antipode in ${\cal P}_2$ and
hence by \thm {\bf
(5.6)} it is orthogonal to all points of ${\cal P}_1$. Using \thm {\bf (8.5)}
and (26) we obtain
\thm {\bf (8.3)} $\bullet$.

The geometric significance of this theorem emerges if in (27) we put $\phi(a,b)
= \theta(a,b)/2$
so that $p(a,b) = \cos^2(\phi(a,b))$ whence our theorem gives:
$$
cos(\phi(z_1,z_2)) = cos(\phi(z_1,z_0))cos(\phi(z_0,z_2)).
\eqno(34)$$
{\it But this may be recognized as the fundamental relation between the
hypotenuse and legs of a
right spherical triangle}~\citeTRIG~. Now consider Figure 3. The two points
$z_1,z_2$ lie on the same
great circle through $a$. Their feet on ${\cal P}_{aa'}$ are $z_1^*$ and
$z_2^*$. Because of $(34)$
the distance from $a$ to $z_1^*$ and from $a$ to $z_2^*$ are such that we can
deduce that $z_1^*$ and
$z_2^*$ {\it must also lie on a single great circle through} $a.$  Thus we have
the general rule
that if we have a great circle $C$ through $a$ lying on one GPS, then the feet
of the points of $C$
on another GPS also form a great circle through $a$. We can
formulate this result as the following very important corollary:

\thm {\bf (8.6)} {\it There is an equivalence relation} ``$\approx$" {\it
between great circles on
GPS's through a given point} $x$ {\it namely:} $C \approx C^*$ {\it if the
points
of} $C$ {\it are the feet on one GPS} ${\cal P}$ {\it of the points of} $C^*$
{\it on another
GPS} $\;{\cal P}^*$. {\it In each class there is one and only one circle from
each GPS through}
$x.$

We call the set consisting  of all points other than the intersection $x$ that
lie on one of the
equivalent circles within a given class a  {\it real cross-section} of ${\cal
S}$. This terminology
is based on the fact that in the case of a real Hilbert space the GPS's are
{\it themselves }
circles, whence the real cross section is the whole of ${\cal S}$ (other than
$x$. Note
that in general we can partition ${\cal S}$ into the disjoint union of real
cross sections plus
$x$.

We shall also need some other corollaries:

\thm {\bf (8.7)} {\it The distance of a point on one GPS to its foot on an
adjacent GPS is the same
for all points at a given distance from the intersection. Moreover the distance
of the foot to the
intersection is the same for all such points.}

Proof:  Let ${\cal P},{\cal Q}$ be two GPS intersecting at $a,$ and let $q \neq
a$ be an
element of ${\cal Q}$ .  Let $m,n$ be the antipodes of $a$ on ${\cal P},{\cal
Q}$ respectively. If
$f$ is the foot on ${\cal P}$ of $q$ then we will show that
$$
d(q,f)/d(q,a) =  d(n,m),
\eqno(35)$$
and the first part of \thm{\bf (8.7)} follows from the fact that the right side
is the same for all
points $q \in {\cal Q}$: Let ${\cal R} = {\cal P}_{mn}$, and $g$ the antipode
of $m$ on ${\cal R}$.
Then $n$ is the foot of q on ${\cal R}$ and
$p(q|n) = 1 - p(q|a) = d^2(q,a)$. Also $p(g|q) = 1 - p(q|f) = d^2(q,f)$ and
$p(g|n) = 1 - p(n|m)
= d^2(n,m)$. Then $(35)$ follows from $p(g|q) = p(g|n)p(n|q)$ given by \thm
{\bf (8.3)}. The
second part then follows from \thm {\bf (8.3)} noting that $p(f|a) = p(a,q)/
p(q,f)$.$\bullet$

With the same notation as in \thm {\bf (8.7)}, let  $u$ be any other point of
${\cal Q}$, $v$  its
foot on ${\cal P}$, and $w$ any other point of ${\cal P}$. Then

\thm {\bf (8.8)}  $ \;\; d(u,v) \leq d(q,w)/d(q,a)$.

Proof: $d(u,v) = d(u,a)d(n,m) \leq d(n,m) = d(q,f)/d(q,a) \leq
d(q,w)/d(q,a)$.$\bullet$

Thus   $d(u,v)$ can be made as small as we like by making $q$ close enough to
some point $w$.

Armed with \thm {\bf (8.3)} we can now prove a very important lemma:  In
the following we assume ${\cal S}$ with $N=3$:

\thm {\bf (8.9)} {\it Refliection of a GPS in a pair of antipodes extends to an
isometry of} ${\cal
S}$ .

Proof: Let ${\cal P}$ be a GPS in a ${\cal S}$, $a,a'$ be a pair of antipodes
of
${\cal P}$, let $a''$ be the unique element such that $a,a',a''$ form a basis
of ${\cal S}$, and
let ${\cal Q} = {\cal P}_{a'a''}$. Then each $z$ in ${\cal Q}$ is orthogonal to
$a$, and we define
the reflection $R$ of ${\cal S}$ to be the reflection of GPS ${\cal P}_{az}$ in
$a,z$ for all $z
\in {\cal Q}$. We contend that this is an isometry of ${\cal S}$.  By \thm{\bf
(8.3)} it suffices to
show that if $x$ is the foot in ${\cal P}_{az}$ of $x^* \in {\cal P}_{az^*}$,
then the image
$Rx$ is the foot in ${\cal P}_{az}$ of the image $Rx^*$. To see this note that
$a,x,z$ define
a great circle $C$ on ${\cal P}_{az}$ through $a$, and $a,x^*,z^*$ define a
great circle $C^*$ on
${\cal P}_{az^*}$ through $a$. By \thm{\bf (8.7)} since  $Rx^*$ is the same
distance
from $a$ as $x^*$, the distance of $Rx^*$ to its foot $f$ on ${\cal P}_{az}$ is
the same as that of
$x^*$ from $x$, and also the distance of $f$ from $a$ is the same as the
distance of $x$ from $a$.
But by \thm{\bf(8.6)} the set of feet of $C^*$ lie on $C,$ and the only point
of $C$ besides
$x$ equidistant from $a$ is the image $Rx$ of $x$, i.e. we must have $f = Rx.$
$\bullet$

\thm{\bf(8.10)} ${\cal S}$ {\it is one-point homogeneous, i.e.\ given any two
points} $x,y$
{\it there is an isometry that maps} $x$ {\it into} $y$.

Proof: We may assume $x \neq y$. Then $x,y$ define a GPS with antipodes $c,c'$
such that $y$ is
the reflection of $x$ in $c,c'$. Hence the assertion follows from
\thm{\bf(8.9)}.

\thm{\bf(8.11)} {\it Any rotation of a GPS about a pair of antipodes can be
extended to an isometry
of} ${\cal S}$.

Proof: Any rotation  of a GPS about a pair of antipodes can be produced by a
pair of successive
reflections. Hence the assertion follows from \thm{\bf(8.9)}.

We are now ready to prove pairwise homogeneity for $N=3$.

 Using \thm{\bf(8.10)} we may first map $a$ to $x$ by an
isometry. It then suffices to show that if $d(a,b) = d(a,y)$ then there is an
isometry holding $a$
fixed that maps $b$ to $y$. Now using \thm{\bf(8.11)} there is an isometry that
rotates ${\cal
P}_{ay}$ so that $y$ is mapped onto the great circle through $a$ containing the
foot $b^*$ of $b$ on
${\cal P}_{ay}$. Thus it suffices to show that for two adjacent GPS there is an
isometry that maps a great circle through the intersection of one into an
equivalent great circle of
the other in the sense of \thm{\bf(8.6)}. But this is accomplished simply as
follows: Let
$a'$ be any fixed element orthogonal to $a$. For each great circle $C$ through
$a$ and $a'$
consider the set of great circles through $a$ equivalent to $C$ in the sense of
\thm{\bf(8.6)}.
Since these circles can be mapped isometrically to a 2-sphere as described in
the proof of
\thm{\bf(8.6)}, it is seen that if $C_1$ is any one of these it will have a
unique reflection
$C_1^*$ in $C$. If we perform this reflection simultaneously for all $C$'s
through $a,a'$, then
since the families are disjoint we will have produced a well-defined reflection
that is also, in
view of \thm{\bf(8.9)}, an isometry. By choosing $a'$ judiciously as the
midpoint of two points
$z,z^*$ antipodal to $a$ on equivalent circles, we will thus have produced the
required isometry.
Combining this with \thm{\bf(6.1)} we have

\thm{\bf(8.12)} {\it Spaces} ${\cal S}$ {\it with} $N=3$ {\it are pairwise
homogeneous, M-convex
metric spaces.}

This is a very important conclusion because of the existence of a theorem due
to Wang ~\citeWANG~
informing us that a pairwise homogeneous, M-convex metric space for $N=3$ must
be one of
the spaces ${\cal R},{\cal C},{\cal Q},{\cal C}ay$ which are the projective
planes over the real,
complex, quaternion, or octonion (Cayley) numbers respectively. The
corresponding  rank of ${\cal S}$
will be 1,2, 4, and 8 respectively. (The reader should not be confused by the
use of the word
``plane". In projective geometry planes have three coordinates.) Note that the
result we have
obtained is much stronger than the result of quantum logic which only informs
us that we
 have {\it some sort} of projective plane. Those that we have obtained are a
very restricted class
that can be  coordinatized by numbers which are {\it almost} fields. The
quaternions and octonions
lack commutativity, and the octonions only obey a restricted form of
associativity (alternative
associativity). From the point of view of projective geometry all of these
planes enjoy a restricted
form of ``transitivity", i.e.\ there is a large but not exhaustive set of
collineations and
correspondingly a large but not exhaustive set of configurations for which
Desargues theorem holds.
In the jargon of projective geometry they are said to be ``Moufang planes".
\vskip.1in
\centerline{{\bf IX. DIMENSIONS $ N > 3$.} }

\indent
We could approach the problem of $N > 3$ by imitating what we did for $N = 3$,
i.e.\ we could
demonstrate that the space is pairwise homogeneous and then invoke Wang's
theorem. However, as in
the case of $N=2$ there is a somewhat more direct approach that takes advantage
of the fact that
for $N > 3$ the representation theorem of projective geometry is very strong.
When combined with
the metric properties we have already derived it will suffice to give us all
that we could get
using Wang's theorem. Moreover this approach enables us to show how coordinates
are actually
introduced and to relate the topology of the coordinates to the d-metric.

 First let us  establish that ${\cal S}$ is a {\it
projective space}. To see this define the term ``{\it line joining $a,b$}" to
mean ${\cal
P}_{ab}$. If $a,b$ are distinct and $z \notin {\cal P}_{ab}$ we construct a
``plane" containing the
three points as we did in \thm{\bf(8.3)} where we constructed the foot of $z$
on ${\cal P}_{ab}$.
Now referring to \thm{\bf(5.4)} one sees that the subspace spanned by $a,b,z$
will obey
all of the axioms for a projective plane~\citeBUM~,~\citeHALL~ provided that
there is at least one more
point not on any of the lines ${\cal P}_{az},{\cal P}_{zb},{\cal P}_{ab}$. This
will be guaranteed
by the connectedness property. This construction applies to all dimensions and
we conclude that
${\cal S}$ is a projective geometry.

Now it is known~\citeBUM~,~\citeHALL~ that for $N > 3$ Desargues theorem holds
throughout the space and
in consequence the space can be coordinatized by a skew-field, i.e.\ all of the
properties of a
field hold except that multiplication need not be commutative. Now it is also
known from a theorem
of Pontriagin ~\citePON~ that the only {\it topological} skew fields are ${\cal
R},{\cal C}$ and
${\cal Q}$. We therefore only have to show that the coordinates respect the
topology of the d-metric
and thereby deduce the same result that would be obtained from Wang's theorem
for $N > 3$, i.e.\
that we must have one of these three projective spaces. The coordinatization
procedure to be
described next is standard~\citeHALL~.

Since every two dimensional subspace is a sphere of the same rank $m$, we can
{\it individually}
coordinatize each such subspace with $m$ real variables  e.g.\ polar latitude
and longitude
variables that are continuous in the $d$-metric. Our task is to  extend this to
the whole three
dimensional space. We proceed in the following  manner~\citeHALL~: For
convenience  write
$xy = P_{xy}$ and call it the ``line joining x and y".  Special reference
points will be indicated
by capital rather than lower case letters. Let O,X,Y be a basis of ${\cal S}$
 with O called the  ``origin" and Y called the ``point at infinity". OX is
called the
X-axis, OY is called the Y-axis, and XY the line at infinity.
 Select some arbitrary point I not on any of the lines OX,OY,XY. For any point
P not on XY let YP
intersect OI at $x$ and XP intersect OI at $y$. Then we coordinatize $P$ by the
pair $(x,y)$. The
line OI then has the equation $y = x$, the points of OX have $y = O$ and of  OY
have $x = O$. (Note
that this is the {\it point} O, not the number zero.) Note also that $O \to
(O,O)$ and $I \to
(I,I)$. Now every point $q$ on XY other than Y is the intersection of a line
from $O$ through some
point of the form $(I,m)$. We write $q \to (m)$. To Y we assign the arbitrary
symbol $(\infty)$.

Now let $(O,b)$ be the intersection on OY of the line through $(m)$ and
$(x,y)$. If this were a
Cartesian plane with the letters indicating real numbers we would have $y = mx
+ b$. In general
since $y$ is determined by $x,m,b$ we write:
$$
y = T(m,x,b).
\eqno(36)$$

The function T with three arguments is called a {\it ternary} operation. We may
now introduce the
convenient definitions:
$$
m\cdot x \equiv T(m,x,O),\quad x + b  = T(I,x,b).
\eqno(37)$$
We must next establish that with these definitions the operations $+$ and
$\cdot$ have
the group and distributivity properties required to define a {\it field} of
numbers. It is a
remarkable fact, however,~\citeBUM~,~\citeHALL~ that the more special cases of
Desargues theorem that
one can identify  as holding in the plane, the closer one can come to a field.
In particular if it is
known that if the plane can be embedded in a higer dimensional space then one
can show (as Desargues
himself did!) that the Desargues theorem holds unrestrictedly. In particular
one can demonstrate that
{}~\citeBUM~,~\citeHALL~
$$
T(m,x,y) = T(I,T(m,x,O),y),
\eqno(38)$$
which means that
$$
T(m,x,y) = m \cdot x + y.
\eqno(39)$$

Projective planes with this property are called ``linear". Moreover one can
show that addition and
multiplication have the group and distributivity properties of a skew-field,
the term ``skew"
meaning  that multiplication need not be
commutative. We now show that this is a {\it topological} field in that $+$ and
$\cdot$ are
continuous operations in the sense of the $d$-metric.

\thm{\bf(9.1)} {\it Addition and multiplication are continuous in the}
$d${\it-metric.}

Proof: Consider a line $L$ joining $(I)$ and $(x,y)$. Let it  intersect $OY$ at
 $U = (O,b)$.
By \thm{\bf(8.8)} the line $L$ comes arbitrarily close to the line $y = x$
joining $(I)$ and O if
$U$ is made close enough to O. But  $y = x + b$ by definition and so we have
proved that $(x,y)$ is
as close as we wish to $(x,x)$ if $b$ is sufficiently close to O in the
topology of OY.  This proves
the continuity of ``$+$". The argument for ``$ \cdot $ " is similar and will be
omitted.$\bullet$

We may now invoke the theorem of Pontriagin ~\citePON~ that informs us that
${\cal R},{\cal C}$ and
${\cal Q}$ are the only topolgical skew-fields. We thus establish that for $N >
3$ we must have one
of these three projective spaces. Note that since all of the listed spaces are
pairwise homogeneous
it follows that

\thm{\bf(9.2)} For all $N$ the space ${\cal S}$ is a pairwise homogeneous,
M-convex, metric
space.

In the quantum logic approach one would at this stage call upon Gleason's
theorem
{}~\citeGLEA~,~\citeVAR~ to deduce (1) using the fact that $p$ is a
frame-function (see $(13)$).
Gleason's theorem has two parts the first of which is deep and difficult, and
the second part of
which is quite simple. The first part tells us that for $N > 2$ the frame
functions on any of the
projective spaces over ${\cal R},{\cal C},{\cal Q},{\cal C}ay$ are {\it
continuous} in the natural
topology of the coordinates. Once this is establshed the easy second part
deduces that the trace
function in (1) is the only permissible form for $p$. Now we see that having
already established
continuity we do not need to invoke the difficult first part of Gleason's
theorem {\it but need only
the second part} to deduce (1). The same remark holds for $N = 3$. Finally in
the case $N=2$ {\it
where Gleason's theorem gives us nothing at all} we do not need it because we
have already
established that $(27)$  gives $p$ in terms of the arc length between points on
the GPS, and in the
particular cases where the rank $m$ is  $1,2,4,8$  the spheres coincide with
{\it projective} lines
over ${\cal R},{\cal C},{\cal Q}$ and ${\cal C}ay$, so that $(27)$ is expressed
by (1).

We have now completed the proof of parts {\bf A,B,C,D,E} of the main theorem
\thm{\bf(4.1)}.
It is remarkable  that the possible models allowed  are precisely the set of
{\it Jordan
algebras}~\citeJWvN~ ! The Jordan algebra axioms codify the algebraic
manipulations that one
performs in conventional quantum mechanics and were introduced with the idea of
discovering
generalizations of quantum mechanics in the early days when it was not clear
that the conventional
model could accommodate relativity. Of course the Jordan axioms do not by
themselves suffice to
deduce (1) and, moreover, they are abstract axioms that are not tied in any
simple way to
experiment. However, the fact that we have found precisely this list of
possibilities is of
considerable interest because the Jordan algebra axioms generalize to the von
Neumann algebras and this suggests possible generalizations of our main
theorem.
\vskip.1in
\centerline{{\bf X. ISOLATING THE CONVENTIONAL MODEL.}}
\vskip.1in
Since the real case is included in the complex case, the problem now arises of
discerning a
physical principle or experiment that isolates the conventional model by
eliminating what we shall
call the {\it exotic} cases, i.e.\  $m > 2$  for $N = 2$,  ${\cal Q}$ for $N
> 2$, and ${\cal C}ay$ for $N = 3$.

Let us begin by looking at $N=2$. Suppose that on a sphere of rank $m$ we
select a basis, i.e.\ a
pair of antipodes. For $m=1$ the equator opposite is a pair of points and for
$m=2$ it is a circle.
The group of isometries leaving the basis fixed in these two non-exotic cases
is a commutative
group, in the first instance the two element group and in the second the group
of rotations about a
fixed axis. But in the exotic cases $m > 2$ the equator is a sphere of rank  at
least two and so
the group of isometries fixing the basis is non-commutative. This same
observation can be
applied for $N > 2$. For there are isometries affecting only two dimensional
subspaces (i.e.\ acting
as the identity on the rest of the space) and in the exotic cases there will be
non-commuting
isometries that leave a basis of the two dimensional subspace invariant. On the
other hand in the
non-exotic cases the isometries that fix a basis are obtained by exponentiating
hermitian operators for which this basis is an eigenbasis. Since
simulataneously diagonalizable
hermitian operators commute the isometries will commute. We thus have the last
part of the main
theorem:

\thm{\bf(4.1 F)} {\it A necessary and sufficient condition for the exclusion of
exotic models
is that isometries fixing all the elements of the same basis commute.}

The question that we must now confront is the physical significance of the
property contained in
\thm{\bf(4.1 F)}. In conventional dynamics time evolution $U(t)$ is associated
with a certain basis, the
eigen-basis of the Hamiltonian $H$ which is invariant under $U$. Any operator
that leaves this basis
invariant is associated with an integral of the motion and commutes with $H$.
Now we see that in the
exotic cases we may have operators that leave that basis invariant but which
fail to commute with
$U$ or (with $H$). It is particularly interesting to consider the quaternionic
case $m = 4$ which is
the only one of the exotic cases that can occur for {\it all} values of $N$. If
we select a basis
which is to be left invariant under time evolution, then there will be a {\it
three-parameter,
non-commutative continuous group} of transformations that leave this basis
invariant. Thus we could
replace the notion of ``time" with a three parameter quantity ${\bf t} =
(t_1,t_2,t_3)$, and
expressions like $e^{-iEt}$ appearing in the evolution of wave functions would
be replaced by
$e^{-iE_1t_1 -jE_2t_2 - kE_3t_3}$ with quaternionic units $i,j,k$. Thus in a
world
with this kind of quantum mechanics both space and time would be three
dimensional and energy like
momentum would be a three-component quantity.

As we remarked earlier any system described by GPS with rank $m > 2$ will also
have additional
elements in the equatorial decomposition and could be identified experimentally
if they existed in
this direct way. For $N > 2$, the exotic cases can also be detected
experimentally by means of the
following construction which we illustrate for $N=3$. Consider Figure 6:

The points $a_1,a_2,a_3$ lie on a GPS ``A" and the points $b_1,b_2,b_3$ lie on
another GPS ``B"
both of which are in the same $N=3$ space. The GPS ${\cal P}_{a_{1}b_{2}}$is
indicated by a line as are the other five GPS's formed by pairings with
distinct subscripts. The
intersection of the indicated GPS's are indicated by  $x,y,z$. It is a
consequence of the
theorem of Pappus in projective geometry over a {\it field} that the three dark
circles will be
collinear. Thus in the conventional model where we have the field of complex
(or real) numbers,
these three will lie on a single GPS. For the exotic cases they will not in
general lie on the same
GPS.
\vskip.1in
\centerline{{\bf XI. CONCLUSIONS.}}
\vskip.1in
Our main theorem \thm{\bf(4.1)} shows that the elementary indistinguishability
properties of mixtures
are sufficient fo imply that the only possible model of quantum mechanics is
given either by (1) or
a few exotic relatives. The results obtained in this way are stronger than
those obtained by
the quantum logic of Birkhoff and von Neumann in that restrictions are obtained
for $N=3$ and even
for $N=2$ where quantum logic and Gleason's theorem give no constraint at all.
We have also described
some direct experimental tests for the low dimensional exotic cases and
indicated the peculiar
dynamical consequences of the exotic variants in arbitrary dimensions.

\vskip.2in

\centerline{{\bf Appendix }}

For conventient reference we here reproduce the
derivation~\citeFIV1~,~\citeFIV2~of the relationship
between $p$ and the $d$-metric in the conventional model and  locally realistic
(hidden variable)
theories.  From (1):
$$
 \sup_z|Tr(\pi(x)\pi(z)) - Tr(\pi(y)\pi(z))| = \sup_z|\langle z|\pi(x)
-\pi(y)|z \rangle|.
\eqno(40)$$
But this is just the largest eigenvalue of $\pi(x) - \pi(y)$. Since the $\pi$'s
 are
projectors:
  $$
(\pi (x) - \pi (y))^3 = (1 - |\langle x|y \rangle|^2)(\pi(x) - \pi(y))
\eqno(41)$$
and one reads off the largest eigenvalue to obtain (25).

If a locally realistic theory is such that there is agreement between its
predictions for various
methods of state preparation~\citeFIV1~ one has a set $\Lambda$ with a measure
$\mu$ such that
$$
p(x|y) = \mu(\Lambda(x)\cap\Lambda(y)),\quad \mu(\Lambda(x)) = 1, \forall x.
\eqno(42)$$
To evaluate (5) we must compute the supremum over $z$ of.
 $|\mu(\Lambda(x)\cap\Lambda(z)) - \mu(\Lambda(y)\cap\Lambda(z))|$. But we note
that the contribution coming from any overlap of $\Lambda(x)$ and $\Lambda(y)$
will cancel. Hence one
can compute the $z$ maximizing the expression  as if the sets are disjoint.
This occurs when
either $z = x$ or $z = y$ and gives $1 - \mu(\Lambda(x)\cap\Lambda(y))$ whence
$$
d(x,y) = 1 - p(x|y).
\eqno(43)$$
Comparing with (25) one sees the incompatibility between quantum mechanics and
locally realistic
theories due to the absence of the square root except in the classical
situation where $p$ assumes
only values $0,1$.
\vfill
\eject
\listrefs
\end